\documentclass[sigconf, nonacm, 11pt]{acmart}

\usepackage{todonotes}
\usepackage{url}

\begin{document}
\begin{titlepage}
   \begin{center}
       \vspace*{1cm}

       \textbf{Technical Report:\\
       Unresolved Challenges and Potential Features in EATXT}

            
       \vspace{1.5cm}

       Contributed by:\\
       \textbf{Dr. Jörg Holtmann}\\
       Independent Researcher\\
       \textbf{Weixing Zhang}\\
       PhD student\\
       Chalmers $|$ University of Gothenburg

       \vspace{1.5cm}
       
       Edited by:\\
       \textbf{Weixing Zhang} \\
       weixing@chalmers.se


       \vfill
            
            
       \vspace{0.8cm}
     
       \includegraphics[width=0.4\textwidth]{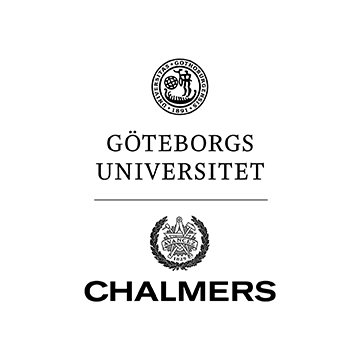}
            
       Computer Science and Engineering\\
       Chalmers $|$ University of Gothenburg\\
       Sweden\\
       2023
            
   \end{center}
\end{titlepage}


%



\pagenumbering{arabic}                
\pagestyle{plain}

\section{Introduction}
The BUMBLE project\footnote{\url{https://itea4.org/project/bumble.html}} was launched to provide an innovative system and software development framework based on blended modeling notations (or languages, such as textual and graphical). 
In this project, the team from the University of Gothenburg (``We'' or ``GU'' in short) conducted technical experiments/implementation on blended modeling\footnote{\url{https://blended-modeling.github.io/}} using EAST-ADL as a case language to explore and improve blended modeling technology. EAST-ADL\footnote{\url{https://www.east-adl.info/Specification.html}} is a language used to describe the architecture of automotive embedded systems. 
EATOP\footnote{\url{https://projects.eclipse.org/projects/modeling.eatop}} is a specialized tool that supports EAST-ADL modeling. Like some of the tools we reported in~\cite{david2023blended}, it supports blended modeling by supporting part of the common notations but does not support the textual notation.
As a start of the work, we used Xtext technology\footnote{\url{https://eclipse.dev/Xtext/index.html}} to develop a textual syntax (and a textual editor that supports it) for EAST-ADL based on the EAST-ADL metamodel. We named this textual syntax and the textual editor \textbf{EATXT}.

We have implemented multiple features in EATXT and have identified some potential features that may be added to EATXT in the future and the challenges involved in implementing them. This document focuses on describing potential advanced features and challenges that have not yet been implemented that can be added to EATXT. This paper does not discuss in detail the features that have been implemented, the use of the software, and the research involved.

\section{Software Development Overview}
As mentioned before, the software we developed is a textual syntax and a textual editor that supports it, we named it EATXT. It supports building, editing, and viewing EAST-ADL models in the form of text, thus filling the gap in textual notation among existing modeling tools that support EAST-ADL.

\begin{figure*}[htb]
	\centering
		\includegraphics[width=\linewidth]{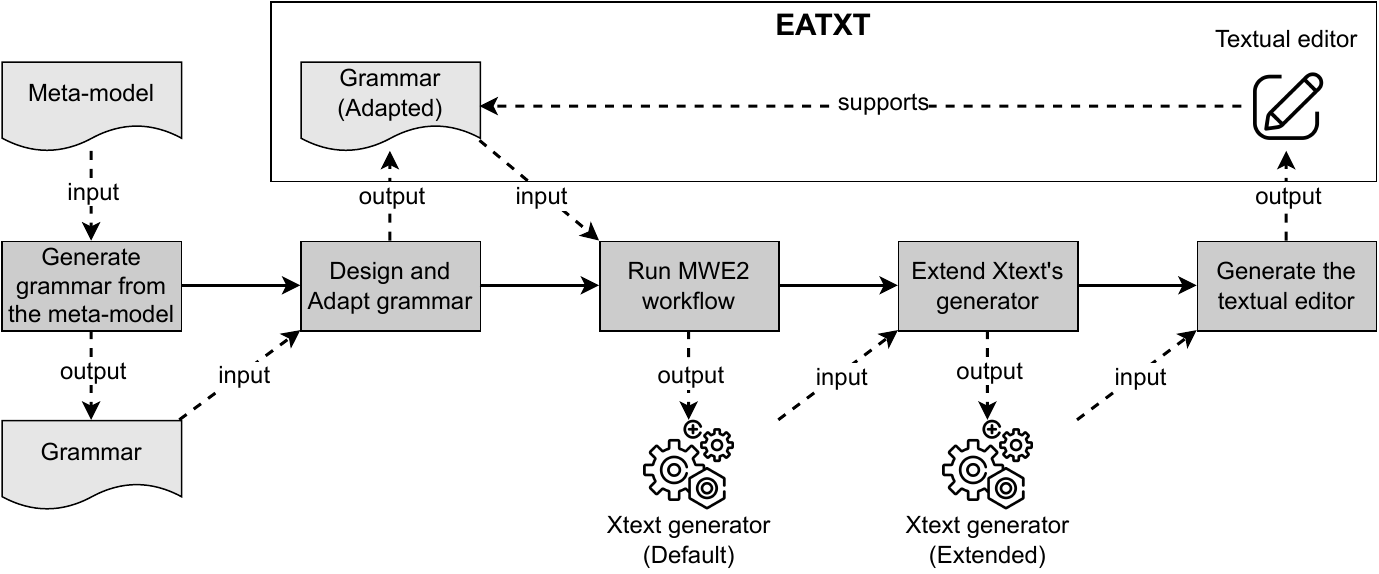}
	\caption{Overview of solution to developing EATXT.}
	\label{fig:solution}
\end{figure*}

Figure~\ref{fig:solution} depicts an overview of our solution to developing EATXT. We first used Xtext to generate the grammar from the EAST-ADL metamodel. We then designed and developed the grammar by adapting the generated grammar, including completing the definition of missing grammar rules, adjusting the grammar style, etc. When the grammar is ready, we run the MWE2 workflow~\footnote{\url{https://wiki.eclipse.org/Modeling_Workflow_Engine_(MWE)}} to generate the Xtext artifacts, which contains the Xtext's generator used to generate the textual editor that supports the grammar. To customize and enhance the features of the textual editor, we extended this Xtext generator. 

\section{Solved Technical Issues}
We have implemented EATXT's grammar by directly adapting the text of the grammar definition, including 1) completing the incomplete grammar rules, 2) creating primitive types, 3) repositioning the attribute \texttt{shortName}, 4) removing redundant elements (such as curly braces), 5) adding symbol ``;'', and 6) supporting empty elements. This adaptation also makes the grammar style tend to be Python style. This style change refers to the work of~\cite{zhang2023pythonstyle}, but retains the use of curly braces to separate code blocks. For information on how to convert a language with a generated grammar into a Python-style language, please refer to~\cite{zhang2023pythonstyle}. We automated this adaptation using the tool GrammarOptimizer~\cite{zhang2023go}.

Moreover, we have added or enhanced some advanced features of the textual editor, including 1) supporting code templates, 2) supporting automatic formatting, 3) supporting content-assist for a new model element with a unique name, and 4) supporting more accurate cross-references. We implemented those features by customizing and extending Xtext's generator capabilities and its implementation which we have introduced in~\cite{holtmann2023Eatxt}. We also implemented conversion (i.e., serialization and deserialization) between textual models (.eatxt model file) and tree-based models (.eaxml model files) on small model instances.

\section{Unsolved Technical Issues}
Unresolved technical issues in EATXT hinder the realization of potential advanced features. Below we will describe the unresolved technical issues by describing these unimplemented advanced features.

\subsection{Blended Modeling of EAST-ADL}
\begin{figure*}[tb]
  \centering
  \includegraphics[scale=0.75]{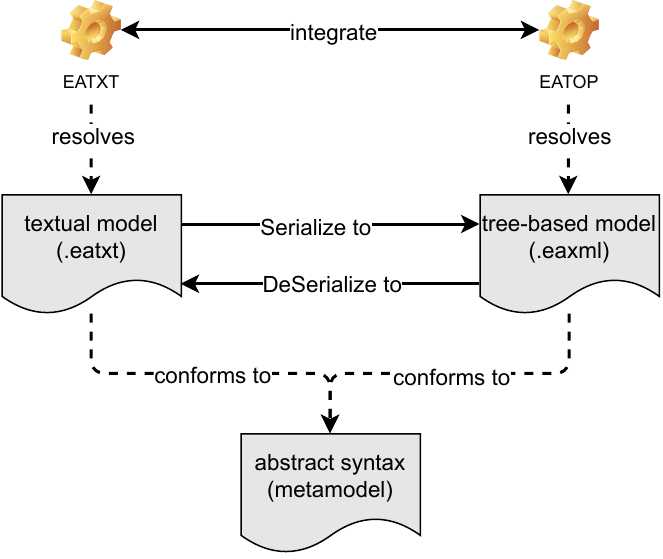}
  \caption{Schematic diagram of EAST-ADL’s blended modeling, including serialization and ensemble..}
  \label{fig:blended_modeling} 
\end{figure*}

Figure~\ref{fig:blended_modeling} depicts how EATXT will work with EATOP to implement blended modeling of EAST-ADL. The current EATXT we implemented can serialize .eatxt model files into .eaxml model files and vice versa, thus achieving blended modeling. The unsolved challenges here are two: 1) This blended modeling feature only works with smaller models, but not larger ones. 2) The current EATXT and EATOP are two independent software, so when performing model conversion, users have to manually import the model file into the software.

\subsection{Flexible Schema Version}

In the second line of the .eaxml file, the field ``xmlns" contains the version of the EAST-ADL metamodel that this file complies with. The version of the metamodel is derived from the EAST-ADL specification it adheres to. Since the initial release of the EAST-ADL specification in 2001, the EAST-ADL metamodel has been evolving. In practice, there are differences between the schemas of consecutive versions, while containing similar essential information. EATOP can tolerate these differences and support multiple versions. However, Xtext is more stringent in this regard. As a result, a textual editor for EAST-ADL developed using Xtext (i.e., EATXT) may not support models adhering to schemas of other EAST-ADL versions. This affects the efficiency of blended modeling in EAST-ADL.

For instance, our industrial partner extended EATOP by developing a graphical view plugin to support the graphical view of models. Although EATOP can accommodate different versions, the graphical view plugin is specific to schema version 2.1.12, and therefore, they model EAST-ADL based on version 2.1.12. Yet, we developed the EATXT based on the latest metamodel version (i.e., 2.2), which does not support loading and resolving models obtained from EATOP due to version mismatch. Consequently, users must manually adapt the models in the .eaxml files before importing them into EATXT. This manual effort adds extra steps in switching between tree-based and textual notation, thereby burdening the efficiency of blended modeling.

To address this issue, EATXT needs to tolerate differences between different versions, enabling users to seamlessly and freely load models from various versions. One possible solution is to have EATXT assess the version of the .eaxml file upon loading. If the version is inconsistent with the EAST-ADL version EATXT is based on, the file should undergo an XSLT transformation to convert it to a format compliant with the version supported by EATXT. By doing so, EATXT ensures compatibility between different EAST-ADL versions and facilitates efficient modeling for users.

\subsection{Error Reporting}

EATXT was built by automatically generating Xtext artifacts from the grammar. It already includes basic error reporting functionality, with error messages displayed in the "Problems" window. However, these error reports are generic and may not provide specific details in certain cases. For example, when a user inputs content that does not match the expected text of the editor, such as typing \texttt{shortName} instead of the required keyword \texttt{EAPackage}, the error message only informs the user of the mismatch without providing further context.

\begin{figure*}[tb]
  \centering
  \includegraphics[width=\linewidth]{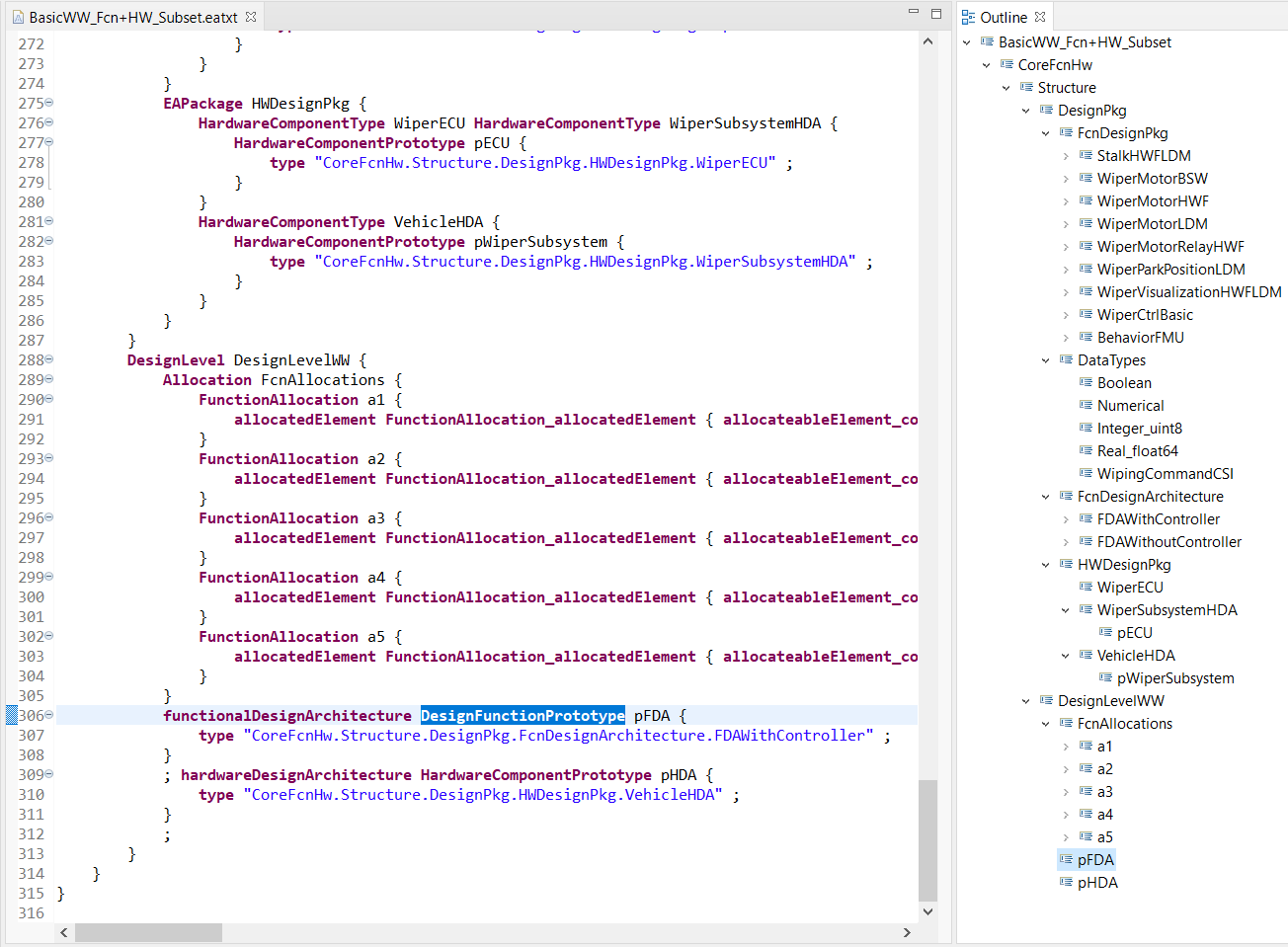}
  \caption{An example of outline in EATXT: \emph{pFDA} stops expanding downwards.}
  \label{fig:eatxt_outlines} 
\end{figure*}

\begin{figure*}[tb]
  \centering
  \includegraphics[width=\linewidth]{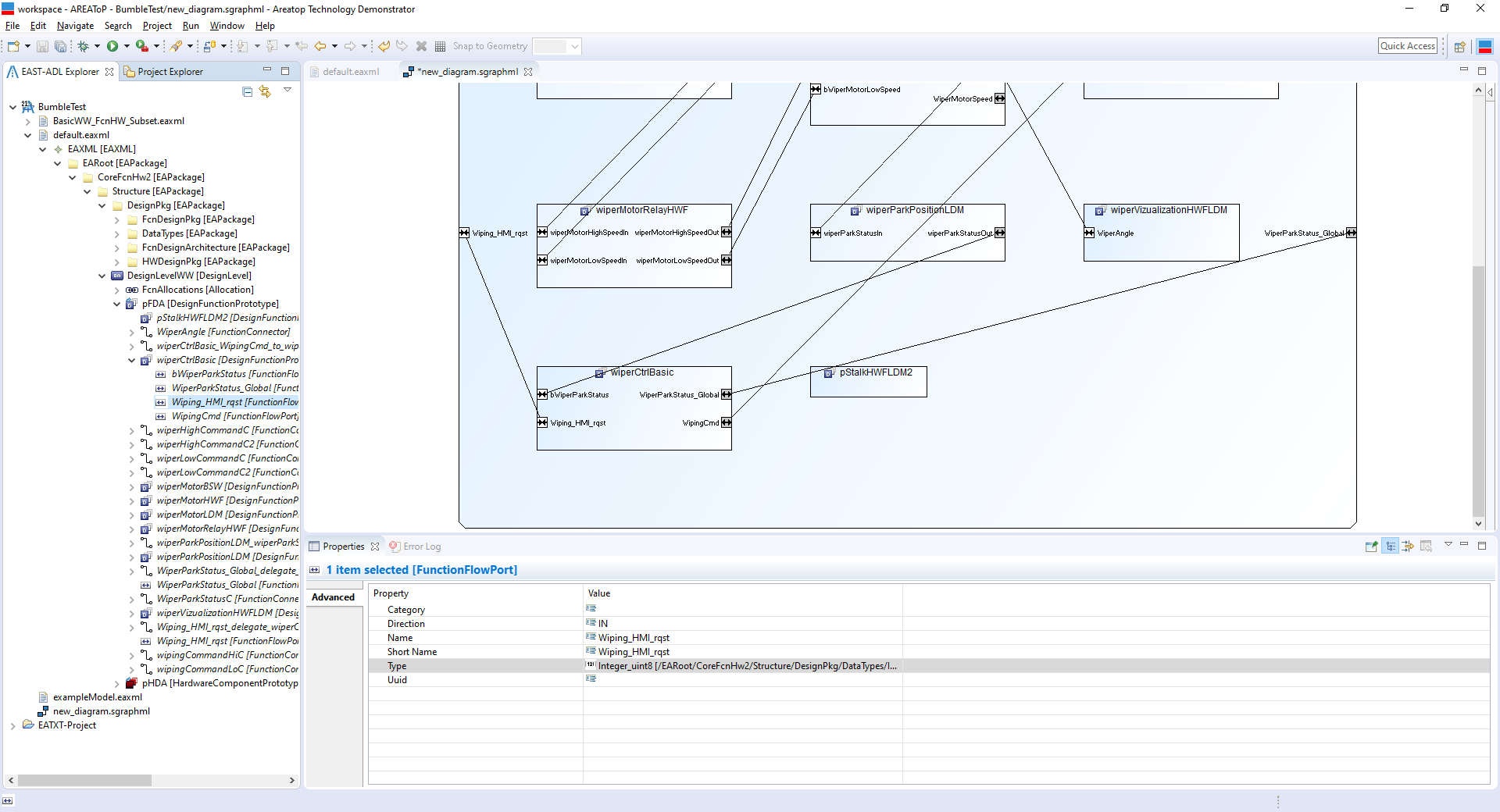}
  \caption{An example of outline in EATOP: \emph{pFDA} continues to expand downwards.}
  \label{fig:eatop_outlines} 
\end{figure*}

Currently, when EATXT attempts to load an incompatible version of an .eaxml file, it rejects the model without providing helpful error messages. We envision that the EATXT in the future will point out the exact issue in such scenarios. For instance, before EATXT loads the .eaxml model file, it could automatically check the version of the EAST-ADL schema that the .eaxml and the EATXT are based on respectively. If there is a version mismatch, EATXT will clearly indicate this in the "Problems" window.

It is important to note that different schema versions may have different elements. For example, in version 2.1.12 of EAST-ADL metamodel, there is an attribute in the class \texttt{HardwareFunctionType} that references to class \texttt{HardwareComponent}, while in version 2.2, such an attribute does not exist. Consequently, a model file (.eaxml) that conforms to version 2.1.12 and contains this attribute would not be loadable by EATXT based on version 2.2. In the envisioned future of EATXT, it would be capable of loading such a file, as described in the previous section, while also providing an accurate error message that the attribute is not necessary for the current version.

\subsection{Precise Outline}

The editor, composed of automatically generated Xtext artifacts, also includes an outline feature, which provides a tree-based representation of model instances, as shown in Figure~\ref{fig:eatxt_outlines}. The current version of EATXT includes this default and generic outline view. However, this outline lacks important details, hindering users' basic understanding of the model structure. For example, Figure~\ref{fig:eatop_outlines} is a screenshot from EATOP and it illustrates an instance of type \texttt{DesignFunctionProtoType} named \texttt{pFDA}, which contains several sub-instances that are cross-referenced to other types. In EATOP's tree-based view, these contents are displayed. However, in the default outline view of the textual editor, the contents below \texttt{pFDA} are hidden, and shown hierarchy stops at \texttt{pFDA}.

\subsection{Integration of EATOP and EATXT}
EATXT facilitates textual modeling of EAST-ADL, with its models capable of being transformed into ".eaxml" model files that can be edited within EATOP, and vice versa. This is the way to implement blended modeling for EAST-ADL. Yet, EATXT is an independent software application separate from EATOP. This signifies that during the blended modeling of EAST-ADL, users are required to run both software applications, necessitating toggling between the two. The challenge here is the integration of one of these applications into the other, with the transformation between ".eatxt" and ".eaxml" files being executed in the background without user intervention. Text-based modeling and tree-based modeling would manifest as two distinct windows within the same software. When a user modifies and saves a model in one window, the software automatically converts the model file, enabling the user to directly open the model in the other window and continue editing in an alternative notation.

\section{Technical Challenges and Future Work}
The challenges we need to address include: realizing schema version identification and tolerance of element differences between different versions of EAST-ADL, customizing error reporting features, rendering deeper hierarchies in outline view, and implementing automatically synchronizing model files of different formats within EATOP. And these will be our future work.

\section{Conclusion}
We report in this document potential and unimplemented advanced features/challenges that could be added to EATXT. Peers working on improving the blended modeling capabilities of the EATOP and those working on developing textual editors with Xtext may benefit from this document.



\bibliographystyle{ACM-Reference-Format}
\bibliography{sample}


\begin{thebibliography}{4}


\ifx \showCODEN    \undefined \def \showCODEN     #1{\unskip}     \fi
\ifx \showDOI      \undefined \def \showDOI       #1{#1}\fi
\ifx \showISBNx    \undefined \def \showISBNx     #1{\unskip}     \fi
\ifx \showISBNxiii \undefined \def \showISBNxiii  #1{\unskip}     \fi
\ifx \showISSN     \undefined \def \showISSN      #1{\unskip}     \fi
\ifx \showLCCN     \undefined \def \showLCCN      #1{\unskip}     \fi
\ifx \shownote     \undefined \def \shownote      #1{#1}          \fi
\ifx \showarticletitle \undefined \def \showarticletitle #1{#1}   \fi
\ifx \showURL      \undefined \def \showURL       {\relax}        \fi
\providecommand\bibfield[2]{#2}
\providecommand\bibinfo[2]{#2}
\providecommand\natexlab[1]{#1}
\providecommand\showeprint[2][]{arXiv:#2}

\bibitem[\protect\citeauthoryear{David, Latifaj, Pietron, Zhang, Ciccozzi, Malavolta, Raschke, Stegh{\"o}fer, and Hebig}{David et~al\mbox{.}}{2023}]%
        {david2023blended}
\bibfield{author}{\bibinfo{person}{Istvan David}, \bibinfo{person}{Malvina Latifaj}, \bibinfo{person}{Jakob Pietron}, \bibinfo{person}{Weixing Zhang}, \bibinfo{person}{Federico Ciccozzi}, \bibinfo{person}{Ivano Malavolta}, \bibinfo{person}{Alexander Raschke}, \bibinfo{person}{Jan-Philipp Stegh{\"o}fer}, {and} \bibinfo{person}{Regina Hebig}.} \bibinfo{year}{2023}\natexlab{}.
\newblock \showarticletitle{Blended modeling in commercial and open-source model-driven software engineering tools: A systematic study}.
\newblock \bibinfo{journal}{\emph{Software and Systems Modeling}} \bibinfo{volume}{22}, \bibinfo{number}{1} (\bibinfo{year}{2023}), \bibinfo{pages}{415--447}.
\newblock


\bibitem[\protect\citeauthoryear{Holtmann, Steghöfer, and Zhang}{Holtmann et~al\mbox{.}}{2023}]%
        {holtmann2023Eatxt}
\bibfield{author}{\bibinfo{person}{Jörg Holtmann}, \bibinfo{person}{Jan-Philipp Steghöfer}, {and} \bibinfo{person}{Weixing Zhang}.} \bibinfo{year}{2023}\natexlab{}.
\newblock \showarticletitle{Exploiting Meta-Model Structures in the Generation of {Xtext} Editors}. In \bibinfo{booktitle}{\emph{11th Intl. Conf. on Model-Based Software and Systems Engineering (MODELSWARD)}}. \bibinfo{pages}{218--225}.
\newblock
\urldef\tempurl%
\url{https://doi.org/10.5220/0011745900003402}
\showDOI{\tempurl}


\bibitem[\protect\citeauthoryear{Zhang, Hebig, Steghöfer, and Holtmann}{Zhang et~al\mbox{.}}{2023a}]%
        {zhang2023pythonstyle}
\bibfield{author}{\bibinfo{person}{Weixing Zhang}, \bibinfo{person}{Regina Hebig}, \bibinfo{person}{Jan-Philipp Steghöfer}, {and} \bibinfo{person}{Jörg Holtmann}.} \bibinfo{year}{2023}\natexlab{a}.
\newblock \showarticletitle{Creating Python-style Domain Specific Languages: A Semi-automated Approach and Intermediate Results}. In \bibinfo{booktitle}{\emph{11th Intl. Conf. on Model-Based Software and Systems Engineering (MODELSWARD)}}. \bibinfo{pages}{210--217}.
\newblock
\urldef\tempurl%
\url{https://doi.org/10.5220/0011744900003402}
\showDOI{\tempurl}
\newblock
\shownote{(in press).}


\bibitem[\protect\citeauthoryear{Zhang, Holtmann, Str{\"u}ber, Hebig, and Steghöfer}{Zhang et~al\mbox{.}}{2023b}]%
        {zhang2023go}
\bibfield{author}{\bibinfo{person}{Weixing Zhang}, \bibinfo{person}{Jörg Holtmann}, \bibinfo{person}{Daniel Str{\"u}ber}, \bibinfo{person}{Regina Hebig}, {and} \bibinfo{person}{Jan-Philipp Steghöfer}.} \bibinfo{year}{2023}\natexlab{b}.
\newblock \bibinfo{title}{Meta-model-based Language Evolution and Rapid Prototyping with Automate Grammar Optimization}.
\newblock
\newblock


\end{thebibliography}

\end{document}